\documentstyle[prd,aps,preprint,amsmath,latexsym,amssymb,floats,epsfig]{revtex}

\def \d{{\mathrm{d}}}

\def \pd{\partial}

\def \iner{\rfloor}
\def \hodge{{}^\star}

\def \tl#1{\overset{\kern 1pt\circ}{#1}}
\def \TL#1{\overset{\kern -3pt \circ}{#1}}
\def \TLL#1{\overset{\kern -7pt \circ}{#1}}

\begin{document}
\title{An elastoplastic theory of dislocations as a physical field theory with torsion}
\author{ Markus Lazar\thanks{Present address: Institute for Theoretical Physics,
     University of Leipzig, Augustusplatz 10, D--04109 Leipzig, 
     Germany, E-mail: lazar@itp.uni-leipzig.de.}}
\address{Max-Planck-Institut f{\"u}r Metallforschung\\ 
     Institut f{\"u}r Physik\\
     D--70569 Stuttgart, Germany} 
\date{\today}    
	
\maketitle
\begin{abstract}
We consider a static theory of dislocations with moment stress 
in an anisotropic or isotropic elastoplastical material as a $T(3)$-gauge theory. 
We obtain Yang-Mills type field equations which express the 
force and the moment equilibrium. 
Additionally, we discuss several constitutive laws between the 
dislocation density and the moment stress.
For a straight screw dislocation, we find the stress field which is modified near
the dislocation core due to the appearance of moment stress.
For the first time, we calculate the localized moment stress, the Nye tensor, 
the elastoplastic energy and the modified Peach-Koehler force of a screw dislocation
in this framework.
Moreover, we discuss the straightforward analogy between a screw 
dislocation and a magnetic vortex. 
The dislocation theory in solids is also considered as a three-dimensional 
effective theory of gravity.
\end{abstract}
\vspace*{5mm}
\noindent
\section{Introduction}
Defects in crystals, e.g., elementary point defects, dislocations, and stacking faults,
play a fundamental role in determining the behaviour and properties of crystalline materials.
In principle, point defects make the crystal viscoelastic, whereas dislocations
cause plasticity.
After plastic bending or twisting a crystal contains dislocations which give 
rise to a lattice curvature. 
The dislocations can directly be observed by the 
help of high resolution electron microscopes.
The crystallographic or topological defects not only influence the mechanical, 
but also the electrical, magnetical and other properties.
All these defects break the symmetry of the ideal crystal (defect-free crystal), as an analogue 
of trivial vacuum, to the real crystal as a nontrivial vacuum.

The traditional description of elastic fields produced by dislocations is based 
on the classical theory of linear elasticity. This approach works for the 
strain and stress field far from the core quite well. However, the 
components of these fields are singular at the dislocation line and
this theory, often applied to practical problems, misses the important feature
of plasticity. 
On the other hand, in conventional plasticity theories no internal length
scale enters the constitutive law and no size effects are predicted.

Therefore, it is quite natural to think of dislocation theory as a 
theory of elastoplasticity.  
In this framework, it is possible to define a characteristic internal length
by the help of a new material constant.
In analogy to the theory of elementary particle physics and gravity, 
we propose a (static) elastoplastic field theory of 
dislocations~(see also~\cite{lazar20}).

A gauge theory of dislocations is formally given in Refs.~\cite{edelen82,edelen83,edelen88} 
but without considering the moment stress. Moreover, their 
gauge Lagrangian of dislocation is not the most general one for an isotropic 
material because it contains only one material constant and has been chosen
in a very special form quadratic in translational gauge field strength.
Recently, Malyshev~\cite{malyshev00} discussed that the gauge Lagrangian
used by Edelen~{\it et al.} does not lead to the correct 
solution for an edge dislocation within a linear approximation.
Additionally, any correct gauge theory of dislocations must give the 
well-established results earlier obtained with the older theory 
of dislocations. For instance, Kadi{\'c} and Edelen~\cite{edelen82,edelen83} 
and Edelen and Lagoudas~\cite{edelen88} find in their gauge theory of 
dislocations that the far field stress of a screw dislocation decays exponentially   
and the near field decay is found to go with $r^{-1}$. Obviously, this is an important 
difference between the gauge theory of dislocations and the classical dislocation theory.
Fortunately, about ten years later Edelen~\cite{edelen96} realized that the 
solutions for a screw and an edge dislocation given 
in Refs.~\cite{edelen82,edelen83,edelen88} are unphysical.

The aim of this paper is to develop a static theory of dislocations
which makes use of the concepts of field strength, excitation, and constitutive 
functions. 
Like Maxwell's field equations, the theory of dislocations consists of two
sets of field equations which are connected by constitutive laws.
This theory is a kind of an axiomatic field theory of dislocations
similar to 
axiomatic Maxwell's theory, which has recently been given by 
Hehl~{\it et al.}~\cite{electro1,obu99}.  
Additionally, this dislocation theory is a 
three-dimensional translation gauge theory~\cite{lazar20} which makes
use of the framework of metric affine gauge theory~(MAG) given
by Hehl~{\it et al.}~\cite{hehl76,hehl80,MAG,Erice95}. 
We discuss as a physical example 
the elastoplastical properties of a screw dislocation as a crystal defect in
full detail. We show that the solution of the gauge field equations 
for a screw dislocation in an infinite medium gives the classical far field
and a modified near field. 

Moreover, we discuss a straightforward analogy between a screw dislocation
and a magnetic vortex in type-II superconductors~\cite{abrikosov,nielsen73}.
Additionally, we discuss the dislocation theory as a three-dimensional theory of gravity.

As formalism we use the calculus of exterior differential forms, 
for our conventions see~\cite{trautman,frankel,marsden,MAG}.

\section{Elasticity theory}
\setcounter{equation}{0}
In classical elasticity theory~(see~\cite{marsden}) the material body is identified with a 
three-dimensional manifold ${\cal M}^3$ which is embedded in the 
three-dimensional Euclidean space ${\Bbb R}^3$. We distinguish between  the 
material or the  final coordinates of ${\cal M}^3$, $a,b,c,\ldots=1,2,3$,  
and the (holonomic) Cartesian coordinates of the reference system  
(defect-free or ideal reference system) ${\Bbb R}^3$, $i,j,k,\ldots=1,2,3$.
A {\it deformation} of ${\Bbb R}^3$ is a mapping ${\boldsymbol\xi}:\ {\Bbb R}^3\rightarrow{\cal M}^3$.
This deformation or distortion one-form is defined by
\begin{align}
\label{dist1}
\vartheta^a=B^a_{\ i}\,\d x^i=\d\xi^a,
\end{align}
and can be identified with the soldering form and the (orthonormal) coframe, respectively.
Here $\d$ denotes the three-dimensional exterior derivative. 
Since
\begin{align}
\d\vartheta^a=\d\d\xi^a=0,
\end{align}
the elastic distortion (\ref{dist1}) is compatible or holonomic and the body manifold is
simply connected.
The compatible distortion one-form~(\ref{dist1}) is invariant under ``rigid'', 
i.e., constant translational transformations,
\begin{align}
\xi^a\longrightarrow \xi^a+ \tau^a,
\end{align}
where $\tau^a$ are constant translations.

Using the orthonormal coframe, the volume three-form is defined by 
\begin{align}
\eta:=\frac{1}{3!}\,\epsilon_{abc}\,\vartheta^a\wedge\vartheta^b\wedge\vartheta^c 
=\frac{1}{3!}\,B\epsilon_{ijk}\,\d x^i\wedge\d x^j\wedge\d x^k ,
\end{align}
with $B\equiv\text{det} (B^a_{\ i})$ and 
$\epsilon_{abc}$ is the Levi-Civita symbol,
$\eta_a:=e_a\iner\eta$, $\eta_{ab}:=e_a\iner e_b\iner\eta$, and 
$\eta_{abc}:=e_a\iner e_b\iner e_c\iner\eta$.
Here $\iner$ denotes the interior product with
\begin{align}
e_a\iner\vartheta^b=B_a^{\ i}B^b_{\ i}=\delta_a^b,\quad e_a=B_a^{\ i}\pd_i,
\end{align}
and $\wedge$ the exterior product ($A\wedge B=A\otimes B-B\otimes A$).
In the following we use the Hodge duality operation~$\hodge$.
For a $p$-form $\alpha$, the dual $(3-p)$-form ($p\le 3$) 
with $\hodge\hodge\alpha=\alpha$ is given by
\begin{align}
\hodge\alpha=\frac{1}{p!}\big( \alpha^{a_1\ldots a_p}\, 
e_{a_p}\wedge\ldots\wedge e_{a_1}\big)\iner\eta.
\end{align}

The {\it Cauchy-Green strain} tensor $G$
is defined as the metric of the final state 
\begin{align}
G=\delta_{ab}\,\vartheta^a\otimes\vartheta^b
=\delta_{ab} B^a_{\ i}B^b_{\ j}\,\d x^i\otimes\d x^j
=g_{ij}\,\d x^i\otimes\d x^j,
\end{align}
where $\delta_{ab}=\text{diag}(+++)$.
We can interpret the strain tensor as a kind of effective field which is formed
out during the deformation.
For an incompressible material, it holds the following condition 
(constraint of incompressibility):
\begin{align}
\text{det} (B^a_{\ i})=\sqrt{\text{det}(g_{ij})}=1.
\end{align}
Finally, the relative strain tensor (Green-Lagrange strain tensor) $E$ 
is given by
\begin{align}
2E=G-1=(g_{ij}-\delta_{ij})\,\d x^i\otimes\d x^j.
\end{align}
It measures the change of the metric between the undeformed and the deformed 
state.

Let us now consider the elastic strain Lagrangian. 
For simplicity we assume linear a 
{\it constitutive law}.
The elastic (anisotropic) Lagrangian is given in terms of the 
potential (strain) energy 
\begin{align}
\label{el_lag1}
{\cal L}_{\mathrm{strain}}=-W\eta.
\end{align}
The {\it potential energy} is given by
\begin{align}
W=\frac{1}{2\cdot 4!}\,\big(C\iner E\otimes E\big) 
=\frac{1}{2}\, C^{ijkl}E_{ij}E_{kl},
\end{align}
where the elasticity tensor~\cite{marsden}, which describes the elastic properties
of the material under consideration, is defined by
\begin{align}
C=C^{ijkl}\,\pd_i\otimes\pd_j\otimes\pd_k\otimes\pd_l ,\qquad
C^{ijkl}=C^{jikl}=C^{ijlk}=C^{klij}.
\end{align}  

The {\it elastic force stress} is  the elastic response quantity pertaining 
to the distortion and is defined by
\begin{align}
\label{stress_el}
\Sigma^{\rm }_a:=\frac{\delta {\cal L}_{\rm strain}}{\delta \vartheta^a}.
\end{align}
We speak of force stresses to distinguish them from the so-called 
moment stress (transmission of moments).
In particular, $\Sigma_a$ is a ${\Bbb R}^3$-valued odd (or axial) differential
form. 
Here $\Sigma^{\rm }_a$ is given by
(see also~\cite{malyshev00,gairola81})
\begin{align}
\label{stress_el2}
\Sigma^{\rm }_a=-\sigma_a^{\ l}\eta_l-W\eta_a,
\end{align} 
with $\sigma^{kl}:=\pd W/\pd E_{kl}=C^{ijkl}E_{ij}$ and 
$\sigma_a^{\ l}=\sigma^{kl}\delta_{ac}B^c_{\ k}$ is the first Piola-Kirchhoff stress 
tensor.
The second term in $\Sigma_a$ is 
due to the variation of the volume three-form $\eta$.
In this way, Eq.~(\ref{stress_el2}) corresponds to
Eshelby's elastic stress tensor~\cite{eshelby51,eshelby75}. 
It is the stress tensor for a compressible medium and
appears in quite natural way in this framework. 

The {\it elastic strain energy} density ${\cal E}_{\rm strain}$ is 
defined as the Hamiltonian
\begin{align}
{\cal E}_{\rm strain}:=-{\cal L}_{\rm strain}
		  =W \eta.
\end{align}

\section{Elastoplasticity--$T(3)$-gauge theory of dislocations}
\setcounter{equation}{0}
In this section, we discuss the theory of elastoplasticity 
as a translational gauge theory ($T(3)$-gauge theory).
We postulate a local $T(3)$ invariance for the field $\xi^a$ 
\begin{align}
\xi^a\longrightarrow \xi^a+\tau^a(x),
\end{align}
where $\tau^a(x)$ are local translations.
If we do it, the invariance of the compatible distortion (\ref{dist1})
is lost under the local transformations.
In order to kill the invariance violating terms, we have to introduce a
compensating gauge potential one-form $\phi^a$, which 
transforms under the local transformations in a suitable form:
\begin{align}
\phi^a\longrightarrow\phi^a-\d\tau^a(x).
\end{align}
The field $\phi^a$ couples in a well determined way to the field $\xi^a$
\begin{align}
\label{dist3}
\vartheta^a=\d \xi^a+\phi^a,
\end{align}
such that the distortion one-form (\ref{dist3}), which is now incompatible,
is invariant under local $T(3)$-transformations.
The coupling in (\ref{dist3}) between the translational gauge potential 
$\phi^a$ and the vector field $\xi^a$ is  
a kind of a translational covariant derivative acting on $\xi^a$~(see also~\cite{MAG,lazar20}).
Accordingly, the incompatible distortion (\ref{dist3}) can be understood as the
(minimal) replacement of the compatible distortion (\ref{dist1}) in 
$T(3)$-gauge theory
\begin{align}
\d\xi^a\longrightarrow \d\xi^a+\phi^a.
\end{align}
The minimal coupling argument leads to the substitution in
the strain energy
\begin{align}
W(\d\xi^a)\longrightarrow W(\d\xi^a,\phi^a).
\end{align}
A translational gauge theory is thus a theory which 
corresponds to the gauge invariance with respect to local displacements 
transformations.

The reason of plasticity are dislocations
and the material gives rise to an elastic response.
The distortion or soldering form (\ref{dist3}) is now anholonomic due 
to $\d\vartheta^a\neq 0$
and the incompatible part is caused by defects. 
The presence of dislocations  makes the final 
crystallographic coordinate system anholonomic and
the body manifold after an incompatible deformation is not simply connected.

If we interpret the dislocation gauge potential $\phi^a$ as the negative
{\it plastic distortion}, we observe
\begin{align}
\d\vartheta^a=-\d\phi^a.
\end{align}
Finally, the {\it total distortion} contains elastic 
and plastic contributions according to
\begin{align}
\d\xi^a=\vartheta^a-\phi^a,
\end{align}
so that the total distortion is compatible and can be written
in terms of the mapping function $\xi^a$.

In Ref.~\cite{lazar20} we have seen that the even (or polar) one-form $\phi^a$ in 
Eq.~(\ref{dist3}) can be 
interpreted as the translational part of the generalized affine connection 
in a Weitzenb{\"o}ck space when the linear connection $\omega^a_{\ b}$ is
globally gauged to zero. 
Such a space carries torsion, but no curvature.

On the other hand, a dislocation is a translational defect which causes
the deviation from the euclidicity of the crystal, sometimes called the inner
geometry. This inner geometry of a crystal with dislocations can be described
as a space with teleparallelism, i.e. a flat space with 
torsion~\cite{kondo52,bilby55,kroener59,kroener60}. In this context, 
a clear physical interpretation of torsion was discovered for the first time. 
The differential geometric notion of torsion was originally introduced by 
{\'E}.~Cartan~\cite{cartan}. In this frame, Cartan had been already found 
that torsion is related to translations.
Additionally, we can identify the 
vector-valued zero-form $\xi^a$ as Trautman's {\it generalized 
Higgs field}~\cite{trautman73}.
In contrast to gravity where the physical meaning of $\xi^a$ is not 
completely clarified~(see~\cite{MAG} and the literature given there), 
we cannot use the ``gauge'' 
condition $\xi^a=x^a$ or $\d\xi^a=0$ in our translation gauge theory of dislocations. 
Consequently, the translational part of the generalized affine connection cannot be identified with the soldering form.
In other words, the translational gauge theory of dislocations is a 
theory where the torsion and the translational part of the 
generalized affine connection play a physical role.


Now we define the translational field strength (torsion) two-form $T^a$, 
in the gauge $\omega^a_{\ b}\equiv 0$, as\footnote{In the 
framework of $T(3)$-gauge theory, the dislocation density is identified 
with the torsion two-form or object of anholonomity in a Weitzenb{\"o}ck space~\cite{lazar20}.
}
\begin{align}
T^a=\d\vartheta^a=\d\phi^a=\frac{1}{2}\, T^a_{\ ij}\,\d x^i\wedge\d x^j.
\end{align}
One obtains the conventional dislocation density tensor $\alpha^a_{\ i}$ from
$T^a$ by means of $\alpha^a_{\ i}=\frac{1}{2}\epsilon_i^{\ jk} T^a_{\ jk}$.
Here, $T^a$ is an even (or polar) two-form with values in ${\Bbb R}^3$.
By taking the exterior derivative one gets the translational Bianchi identity
\begin{align}
\label{bianchi}
\d T^a=0.
\end{align}
Physically, Eq.~(\ref{bianchi}) means that dislocations cannot end inside 
the body~\cite{kosevich}.
A characteristic quantity  which expresses a fundamental property of 
dislocations is the Burgers vector.
The Burgers vector\footnote{The non-vanishing of the integral
(\ref{burger}) has topological reasons, see section~\ref{topo}.}
is defined by integrating around a closed path $\gamma$ (Burgers circuit)
encircle a dislocation
\begin{align}
\label{burger}
b^a=\oint_\gamma \vartheta^a=\int_{S} T^a,
\end{align}
where $S$ is any smooth surface with boundary $\gamma=\pd S$.
Thus, the dislocation shows itself by a closure failure (Burgers vector),
i.e. a translational misfit.
This means that the closed parallelogram of the ideal crystal does not 
close in the dislocated crystal.
For a distribution of dislocations we have to interpret $b^a$ in 
Eq.~(\ref{burger}) as the sum of the Burgers vectors of all dislocations
which pierce through the surface $S$ (dislocation density flux 
through the surface~$S$).

To complete the field theory of dislocations, we have to define the 
excitation with respect to the dislocation density. 
We make the most general Yang-Mills-type ansatz 
\begin{align}
\label{L_disl}
{\cal L}_{\rm disl}=-\frac{1}{2}\,T^a\wedge H_a.
\end{align}
Here the {\it moment stress} one-form $H_a$ is defined 
by~(see~\cite{lazar20,gairola81,trzesowski90,trzesowski93,kroener93,kroener95})
\begin{align}
H_a:=-\frac{\pd{\cal L}_{\rm disl}}{\pd T^a}
\end{align}
as an odd (or axial) ${\Bbb R}^3$-valued form. It is sometimes called 
couple stress.
The moment stress is the {\it elastoplastic excitation} with respect to
$T^a$. In other words, that at all positions where the dislocation density is 
non-vanishing, moment stresses occur.
Hence, dislocation theory is a couple or moment stress theory (see also~\cite{kroener68}).
The physical meaning of the couple stress $H_a=H_{ai}\d x^i$ is: 
the components $H^l_{\ l}$ describe twisting moments and the other 
components bending moments~\cite{kroener63,hehl65}.

In order to give concrete expressions for the excitation, 
we have to specify the {\it constitutive relation} between the field strength
$T^a$ and the excitation $H_a$. 
We choose a linear constitutive law for an anisotropic material 
as\footnote{A similar constitutive law between moment stress and 
dislocation density 
was discussed in Refs.~\cite{kroener63,hehl65}.}
\begin{align}
\label{const_ani_lin1}
H_a=\frac{1}{2}\,\hodge\Big(\kappa_{aij}^{\ \ \ \, bkl}\, T_{bkl}\,\d x^i\wedge\d x^j\Big),
\end{align}
where $\kappa_{aij}^{\ \ \ \, bkl}(x)$ 
are constitutive functions that are characteristic for a crystal
with dislocations. These constitutive functions are necessary because 
the elasticity tensor says nothing about the behaviour in the core of 
dislocations (plastic region).
They have the symmetries
\begin{align}
\kappa^{aijbkl}=\kappa^{bklaij}=-\kappa^{ajibkl}=-\kappa^{aijblk}.
\end{align}  
For an isotropic material the most general constitutive law is given by
\begin{align}
\label{const_iso}
H_a=\,\hodge\!\sum_{I=1}^{3}a_{I}\,^{(I)}T_a.
\end{align}
Here $a_1$, $a_2$ and $a_3$ are new material constants for 
a dislocated material.
We use the decomposition of the torsion 
$T^a=\,^{(1)}T^a+\,^{(2)}T^a+\,^{(3)}T^a$ into its $SO(3)$-irreducible 
pieces. These three pieces $^{(I)}T_a$ are (see also~\cite{MAG})
\begin{alignat}{2}
\label{tentor}
^{(1)}T^a&:=T^a-\,^{(2)}T^a-\,^{(3)}T^a
&&\qquad \text{(tentor)},\\
\label{trator}
^{(2)}T^a&:=\frac{1}{2}\vartheta^a\wedge\big(e_b\iner T^b\big)
&&\qquad\text{(trator)},\\
\label{axitor}
^{(3)}T_a&:=\frac{1}{3}e_a\iner\big(\vartheta^b\wedge T_b\big)
&&\qquad\text{(axitor).}
\end{alignat}
The tentor is the torsion corresponding to the Young tableau $(2,1)$ 
minus traces. The trator contains the trace terms of the Young tableau
$(2,1)$
and the axitor corresponds to the Young tableau $(1,1,1)$ (for group-theoretical
notations see, e.g.,~Refs.~\cite{Hamer,BR}). 

The stress two-form of dislocations is defined by
\begin{align}
h_a:=\frac{\pd{\cal L}_{\rm disl}}{\pd\vartheta^a}
=e_a\iner {\cal L}_{\rm disl}+\big(e_a\iner T^b\big)\wedge H_b
=\frac{1}{2}
\Big[\big(e_a\iner T^b\big)\wedge H_b-\big(e_a\iner H_b\big)T^b\Big].
\end{align}
It is an odd (or axial) vector-valued form. This stress form is called
the {\it Maxwell stress two-form of dislocations}.
Here, $h_a$ is a kind of interaction stress between dislocations which
reflects nonlinearity and universality of interactions of the dislocation 
theory. It is expressed in terms of dislocation
density and moment stress. 
Thus, $h_a$ describes higher order stresses in the core region. 
A similar interaction stress of dislocations is
discussed in Refs.~\cite{maugin,stumpf}. 

The definition of the pure {\it dislocation energy} as the 
Hamiltonian is given by
\begin{align}
\label{H_disl}
{\cal E}_{\rm core}:=-{\cal L}_{\rm disl}
		   =\frac{1}{2}\,T^a\wedge H_a.
\end{align}
More physically, we can interpret ${\cal E}_{\rm core}$ as the static 
dislocation core energy density. 


In order to take boundary conditions into account we use a so-called
null Lagrangian~\cite{edelen88}:
\begin{align}
\label{null-L}
{\cal L}_{\rm bg}=\d\left(\sigma^{\rm bg}_a \xi^a\right)	
	         = \sigma^{\rm bg}_a\wedge\d\xi^a
	         \longrightarrow \sigma^{\rm bg}_a\wedge\vartheta^a. 
\end{align}
A null Lagrangian does not change the Euler-Lagrange equations in
classical elasticity (force equilibrium) because 
the background stress $\sigma^{\rm bg}_a$ is required to satisfy the relation
$\d\sigma^{\rm bg}_a=0$. 
After minimal replacement in Eq.~(\ref{null-L}), 
the Lagrangian ${\cal L}_{\rm bg}$ will make
contributions to the Euler-Lagrange equations of elastoplasticity
(see also Ref.~\cite{edelen96}).

The variation of the total Lagrangian
\begin{align}
\label{L}
{\cal L}={\cal L}_{\rm disl}+{\cal L}_{\rm strain}+{\cal L}_{\rm bg}
\end{align}
with respect to $\xi^a$ and $\phi^a$  gives the following field 
equation in the elastoplastical theory of dislocations in an infinite 
medium 
\begin{alignat}{2}
\label{eq}
&\frac{\delta {\cal L}}{\delta \xi^a}\equiv\d\Sigma^{\rm }_a+\d h_a=0
&&\qquad\text{(force equilibrium)},\\
\label{YM-fe}
&\frac{\delta {\cal L}}{\delta \phi^a}\equiv\d H_a-h_a=\widehat\Sigma^{\rm }_a
&&\qquad\text{(moment equilibrium)},
\end{alignat}
where the effective stress two-form, 
$\widehat\Sigma_a:=\Sigma_a+\sigma^{\rm bg}_a$, is the driving force stress
for the moment stress in Eq.~(\ref{YM-fe}).
Let us note that in the framework of MAG, Eq.~(\ref{YM-fe}) is the (first) 
gauge field equation and (\ref{eq}) is the matter field equation (see~\cite{MAG}).
They are Yang-Mills type field equations of the translational gauge theory. 

In order to complete the framework of elastoplastic field theory, 
we define the elastoplastic forces as field strength $\times$ stress.
We introduce the {\it elastic material force} density by the help of the  
material stress two-form via
\begin{align}
\label{force1}
f^{\rm el}_a=\big(e_a\iner T^b\big)\wedge\Sigma_b.
\end{align}
This force contains the contributions due to the eigenstress of 
dislocations (Peach-Koehler force~\cite{peach}).
The pure dislocation force is given by means of the stress two-form of 
dislocations as
\begin{align}
\label{force2}
f^{\rm disl}_a=\big(e_a\iner T^b\big)\wedge h_b.
\end{align}
This force (\ref{force2}) 
characterizes the interaction between dislocations near the dislocation core.

\section{What would be a good choice for the moment stress?}
\setcounter{equation}{0}
In this section we want to discuss different choices for the material
constants $a_1$, $a_2$, and $a_3$. Additionally, we consider the corresponding
equations for the moment equilibrium.

For simplicity, we use the weak field approximation (linearization)
\begin{align}
\xi^a=\delta_i^a x^i+u^a,\qquad  \vartheta^a=\big(\delta^a_{i}+\beta^a_{\ i}\big)\d x^i,
\end{align}
where $u^a$ is the displacement field and $\beta^a_{\ i}$ is the linear 
distortion tensor.
We note that the dislocation self-interaction stress tensor $h_a$ 
is of higher order in the Burgers vector. Thus, we neglect it. 
Here we assume a linear asymmetric (Piola-Kirchhoff) force stress~\cite{schaefer}
\begin{align}
\label{ss-gen}
\sigma_{a}=
2\mu\left(\beta_{(ai)}+\frac{\nu}{1-2\nu}\,\delta_{ai} \beta^k_{\ k}
+c_1\,\beta_{[ai]}\right)\d x^i,
\end{align}
where $\mu$ is the shear modulus and $\nu$ Poisson's ratio. The constant
$c_1$ characterizes the antisymmetric force stress.
Then the equation of the moment equilibrium reads
\begin{align}
\d H_a=\widehat\Sigma_a,
\end{align}
with $\hodge\widehat\Sigma_a=-\widehat\sigma_a$.

The simplest choice is $a_1=a_2=a_3$. It yields the moment stress as
\begin{align}
\label{moment_1}
H_a= H_{ak}\,\d x^k
   =\frac{a_1}{2}\,  T_{aij}\epsilon^{ij}_{\ \, k}\,\d x^k
   =a_1\, \alpha_{ai}\,\d x^i,
\end{align}
with $T^a_{\ ij}=\epsilon_{ijk} \alpha^{ak}$.
This is what Edelen did~\cite{edelen96} in his gauge theory of dislocations.
Accordingly, in this connection we call this choice ``Edelen choice''.
Eventually, we obtain the field equation 
(see, e.g., Eq.~(2.6) in Ref.~\cite{edelen96})
\begin{align}
\label{FE-lin2}
\Delta\beta_{ai}-\pd_i\pd^j\beta_{aj}=\frac{1}{a_1}\,\widehat\sigma_{ai},
\end{align}
where $\Delta$ denotes the Laplace operator 
and $\widehat\sigma_{ai}=\sigma_{ai}-\sigma^{\rm bg}_{ai}$.
Now we use the decomposition $\beta_{ai}=\beta_{(ai)}+\beta_{[ai]}$ and
\begin{align}
\label{ss-sy}
\beta_{(ai)}&=
\frac{1}{2\mu}\left(\sigma_{(ai)}-\frac{\nu}{1+\nu}\,\delta_{ai} \sigma^k_{\ k}\right),\\
\label{ss-asy}
\beta_{[ai]}&=\frac{1}{2\mu c_1}\,\sigma_{[ai]}.
\end{align}
We find from Edelen's field equation~(\ref{FE-lin2}) the equations for the 
symmetric and antisymmetric force stress as 
\begin{align}
\label{FE-lin2-sy}
&\Delta\sigma_{(ai)}-\pd_i\pd^j\sigma_{(aj)}+
\frac{\nu}{1+\nu}\left(\pd_a\pd_i-\delta_{ai}\Delta\right)\sigma^k_{\ k}
=\kappa^2\,\widehat\sigma_{(ai)},
\qquad\kappa^2=\frac{2\mu}{a_1},\\
\label{Fe-lin-2asy}
&\Delta\sigma_{[ai]}-\pd_i\pd^j\sigma_{[aj]}
=c_1\, \kappa^2\,\widehat\sigma_{[ai]}.
\end{align}
Therefore, the Edelen choice in combination with Eq.~(\ref{ss-gen}) yields
to equations for symmetric and antisymmetric force stresses.
If we require a symmetrical force stress by setting $c_1=0$ in Eq.~(\ref{ss-gen}) 
and use the force equilibrium condition $\pd^i\sigma_{ai}=0$ as 
``gauge condition'', we obtain from Eq.~(\ref{FE-lin2-sy}) the field 
equation
\begin{align}
\label{FE-lin3}
\Delta\sigma_{(ai)}+
\frac{\nu}{1+\nu}\left(\pd_a\pd_i-\delta_{ai}\Delta\right)\sigma^k_{\ k}
=\kappa^2\,\widehat\sigma_{(ai)}.
\end{align}

Another choice could be  $a_1=a_2=0$. Then the axitor, 
\begin{align}
\label{axitor2}
^{(3)}T_{aij}=\frac{1}{3}\left(T_{aij}+T_{ija}+T_{jai}\right),
\end{align}
defines the moment stress as
\begin{align}
\label{moment_2}
H_a=\frac{a_3}{6}\,
\left(T_{aij}+T_{ija}+T_{jai}\right)
\epsilon^{ij}_{\ \, k}\,\d x^k
=\frac{a_3}{3}\,\delta_{ai}\alpha^{k}_{\ k}\,\d x^i.
\end{align}
By the help of the axitor,
the field equation for the distortion field is given by
\begin{align}
\label{fe_axitor1}
\Delta\beta_{[ai]}-\pd_i\pd^j\beta_{[aj]}+\pd_a\pd^j\beta_{[ij]}
=\frac{3}{2 a_3}\,\widehat\sigma_{ai}.
\end{align}
If we use Eq.~(\ref{ss-asy}) and $\pd^i\sigma_{ai}=0$, we find the equation 
\begin{align}
\label{fe_axitor2}
\Delta\sigma_{[ai]}=c_1\,\tilde\kappa^2\,\widehat\sigma_{[ai]},\qquad 
\tilde\kappa^2=\frac{3\mu}{a_3},
\end{align}
and $\widehat\sigma_{(ai)}=0$.
It follows from Eq.~(\ref{fe_axitor2}) that this 
choice of the moment stress requires a pure antisymmetric force stress.

Now we discuss the choice $a_2=0$ and $a_3=-\frac{a_1}{2}$.
Then the moment stress is given in terms of the contortion tensor 
according to
\begin{align}
\label{moment_4a}
H_a=\frac{a_1}{4}\, 
\big(T_{aij}-T_{ija}-T_{jai}\big)\epsilon^{ij}_{\ \, k}\,\d x^k.
\end{align}
If we use $T^a_{\ ij}=\epsilon_{ijk} \alpha^{ak}$, we find the relation between
the moment stress tensor $H_{ai}$ and the Nye\cite{Nye} 
tensor $\kappa_{ai}=\alpha_{ia}-\frac{1}{2}\delta_{ai}\alpha^k_{\ k}$ as
\begin{align}
\label{moment_4b}
H_a=a_1\left(\alpha_{ai}-\frac{1}{2}\,\delta_{ai}\alpha^k_{\ k}\right)\,\d x^i
\equiv a_1\, \kappa_{ia}\,\d x^i.
\end{align}
Hence, the moment stress tensor $H_{ai}$ given in this choice is proportional
to the transpose of the Nye tensor.
The field equation by means of this moment stress is given by
\begin{align}
\label{FE-a12}
\Delta\beta_{(ai)}-\pd_i\pd^j\beta_{(aj)}-\pd_a\pd^j\beta_{[ij]}=\frac{1}{a_1}\,\widehat\sigma_{ai}.
\end{align}
Again, we use the Eqs.~(\ref{ss-sy}) and (\ref{ss-asy}) and find the 
equations for the symmetric and antisymmetric force stress
\begin{align}
&\Delta\sigma_{(ai)}-\pd_i\pd^j\sigma_{(aj)}+
\frac{\nu}{1+\nu}\left(\pd_a\pd_i-\delta_{ai}\Delta\right)\sigma^k_{\ k}
=\kappa^2\,\widehat\sigma_{(ai)},\\
&-\pd_a\pd^j\sigma_{[ij]}
=c_1\,\kappa^2\,\widehat\sigma_{[ai]}.
\end{align}
If we put $c_1=0$ (symmetrical force stress) and use $\pd^i\sigma_{ai}=0$, 
we obtain 
\begin{align}
\label{FE-lin4}
\Delta\sigma_{(ai)}+
\frac{\nu}{1+\nu}\left(\pd_a\pd_i-\delta_{ai}\Delta\right)\sigma^k_{\ k}
=\kappa^2\,\widehat\sigma_{(ai)},
\end{align}
which agrees with Eq.~(\ref{FE-lin3}).

Another interesting choice seems to be $a_2=-a_1$ and $a_3=-\frac{a_1}{2}$.\footnote{This 
choice of parameters is called the ``Einstein choice'' and 
can be obtained from the condition that the gauge Lagrangian ${\cal L}_{\rm disl}$ 
has to be invariant under local $SO(3)$-transformations
in order to obtain the teleparallel version of the Hilbert-Einstein 
Lagrangian in three dimensions\cite{katanaev92}.}
Then we use the axitor~(\ref{axitor2}), 
the trator, 
\begin{align}
^{(2)}T_{aij}=\frac{1}{2}\left(\delta_{ai}T^l_{\ lj}
                         +\delta_{aj}T^l_{\ il}\right),
\end{align}
and the tentor, 
\begin{align}
^{(1)}T_{aij}=T_{aij}-\,^{(2)}T_{aij}-\,^{(3)}T_{aij}.
\end{align}
Consequently, the moment stress is given by
\begin{align}
\label{moment_4}
H_a=\frac{a_1}{4}\, 
\big(T_{aij}-T_{ija}-T_{jai}
-2\delta_{ai}T^l_{\ lj}-2\delta_{aj}T^l_{\ il}\big)
\epsilon^{ij}_{\ \, k}\,\d x^k.
\end{align}
We find the remarkable relationship between
the moment stress tensor $H_{ai}$ and the Nye tensor $\kappa_{ai}$ as
\begin{align}
\label{moment5}
H_{ai}= a_1\left(\alpha_{ia}-\frac{1}{2}\,\delta_{ai}\alpha^k_{\ k}\right)
	\equiv a_1\kappa_{ai}.
\end{align}
Thus, this moment stress tensor $H_{ai}$ is proportional to the
Nye tensor.
Eventually, we obtain the field equation for the distortion $ \beta_{(ai)}$ as
\begin{align}
\label{fe-E}
a_1\left\{\Delta \beta_{(ai)}-(\pd_i\pd^k \beta_{(ak)}+\pd_a\pd^k \beta_{(ik)})
+\delta_{ai}\pd^k\pd^l \beta_{(kl)}+\pd_a\pd_i \beta^k_{\ k}
-\delta_{ai}\Delta \beta^k_{\ k}\right\}=\widehat\sigma_{(ij)}.
\end{align}
Because the l.h.s. of Eq.~(\ref{fe-E}) is equivalent to
$(\text{inc}\, \boldsymbol{\beta})_{(ij)}$, 
Eq.~(\ref{fe-E}) is the proper gauge theoretical formulation of 
Kr{\"o}ner's incompatibility equation (see, e.g., Eq.~(II.21) in Ref.~{\cite{kroener58}).
Kr{\"o}ner's incompatibility tensor is replaced by the 
effective stress tensor on the r.h.s. of this equation.
Additionally, we rewrite the field equation in terms of the force stress as
\begin{align}
\label{beltrami}
\Delta\sigma_{(ai)}+\frac{1}{1+\nu}\big(\pd_a\pd_i
-\delta_{ai}\Delta\big)\sigma^k_{\ k}=\kappa^2\, \widehat\sigma_{(ai)}.
\end{align}
For $\widehat\sigma_{(ai)}=0$ this equation is the Beltrami equation.
Let us emphasize that the factor $1/(1+\nu)$ in Eq.~(\ref{beltrami}) 
differs from the factor $\nu/(1+\nu)$ in Eqs.~(\ref{FE-lin3}) and 
(\ref{FE-lin4}). Another interesting point is that by means of the 
``Einstein choice'' the force stress is symmetric in quite natural way. 
We do not have to make any assumption with respect to the material constant
$c_1$.
Therefore, in order to investigate dislocations with symmetric force
stress the ``Einstein choice'' of the constants $a_1$, $a_2$, and $a_3$ 
is favourable and will be used in the following.

\section{A straight screw dislocation in linear approximation}
\setcounter{equation}{0}
\subsection{Field equation and stress field}
For simplicity, we consider a straight screw dislocation in an isotropic 
and incompressible medium ($B=1$) in linear approximation. 
For a straight screw dislocation, the Burgers 
vector and the dislocation line are parallel.
In this case, the problem has {\it cylindrical} symmetry.

In the framework of the elastoplastic field theory,
we require that the modified stress field of a screw dislocation 
has the following properties:
(i)~the stress field should have no singularity at $r=0$,
(ii)~the far field stress ought to be the stress field of a 
Volterra dislocation $\sigma_{ij}^{\rm bg}$ which satisfies
the condition $\pd^j\sigma^{\rm bg}_{ij}=0$. 
Thus the condition (ii) is a boundary condition
for the stress of a dislocation in the field theory of elastoplasticity.
We choose the dislocation line and Burgers
vector in the $z$-axis of a Cartesian coordinate system. 
Then the background stress is given by the elastic stress of a 
Volterra screw dislocation~\cite{HL}
\begin{align}
\sigma^{\rm bg}_{xz}=\sigma^{\rm bg}_{zx}=
-\frac{\mu b}{2\pi}\,\frac{y}{r^2},
\quad
\sigma^{\rm bg}_{yz}=\sigma^{\rm bg}_{zy}=
\frac{\mu b}{2\pi}\,\frac{x}{r^2},
\end{align}
where $r^2=x^2+y^2$.
Obviously, these stress fields are singular at the dislocation line.

We turn to Eq.~(\ref{beltrami}) and put $\sigma^k_{\ k}=0$.
Then the field equation for the force stress of a linear screw dislocation is
given by the following inhomogeneous Helmholtz equation, 
\begin{align}
\label{stress-fe-screw}
\left(1-\kappa^{-2}\Delta\right)\sigma_{(ij)}=\sigma^{\rm bg}_{(ij)}.
\end{align} 
Thus, for the elastic strain fields,
\begin{align}
\left(1-\kappa^{-2}\Delta\right)E_{ij}=E^{\rm bg}_{ij}.
\end{align} 
If we put $\sigma^k_{\ k}=0$ in Eqs.~(\ref{FE-lin3}) and (\ref{FE-lin4}), 
we also obtain Eq.~(\ref{stress-fe-screw}). 

Now we seek for a cylindrically symmetric (string-like) solution of a screw
dislocation.
One finds the solution for the distortion field
\begin{align}
\beta_{zx}=-\frac{y}{r^2}\left(\frac{b}{2\pi}+C_1 r K_1(\kappa r)\right),\qquad
\beta_{zy}=\frac{x}{r^2}\left(\frac{b}{2\pi}+C_1 r K_1(\kappa r)\right),
\end{align}
where $K_1$ is the modified Bessel function of the second kind of order one.
This solution is similar to the potential of a magnetic vortex 
(Abrikosov-Nielsen-Olesen string) for a constant Higgs field~\cite{abrikosov,nielsen73}.  
The constant of integration $C_1$ is determined
form the condition that the distortion $\beta_{zx}$ and $\beta_{zy}$ 
vanish at $r=0$
($\kappa r\ll 1$ with $C_1 K_1(\kappa r)\approx C_1 \frac{1}{\kappa r}$)
as
\begin{align}
\label{C1}
C_1=-\frac{b\kappa}{2\pi}.
\end{align}
With (\ref{C1}) we obtain for the distortion of a screw dislocation  
\begin{align}
\beta^{z}=\frac{b}{2\pi r^2}\big(1-\kappa r K_1(\kappa r)\big)
\left(x\,\d y-y\,\d x\right).
\end{align}
The distortion one-form can be expressed in cylindrical coordinates
as
\begin{align}
\label{disl-pot}
\beta^{z}=\frac{b}{2\pi}\big(1-\kappa r K_1(\kappa r)\big)\,\d\varphi.
\end{align}
The deformation  (\ref{disl-pot}) around the screw dislocation is a pure shear.
The effective Burgers vector is given by
\begin{align}
b^z(r)=\oint_\gamma\beta^z=b\big\{1-\kappa r K_1(\kappa r)\big\}.
\end{align}
This effective Burgers vector $b^z(r)$ differs from the constant Burgers
vector $b$ in the region from $r=0$ up to $r=6/\kappa$ 
($\{1-\kappa r K_1(\kappa r)\}|_{\kappa r=6}=0.992$) 
because the distortion field is modified due to the moment stress~(see~Fig.~\ref{fig:burger}).
\begin{figure}[h]\unitlength1cm
\centerline{
\begin{picture}(9,6)
\put(0.0,0.2){\epsfig{file=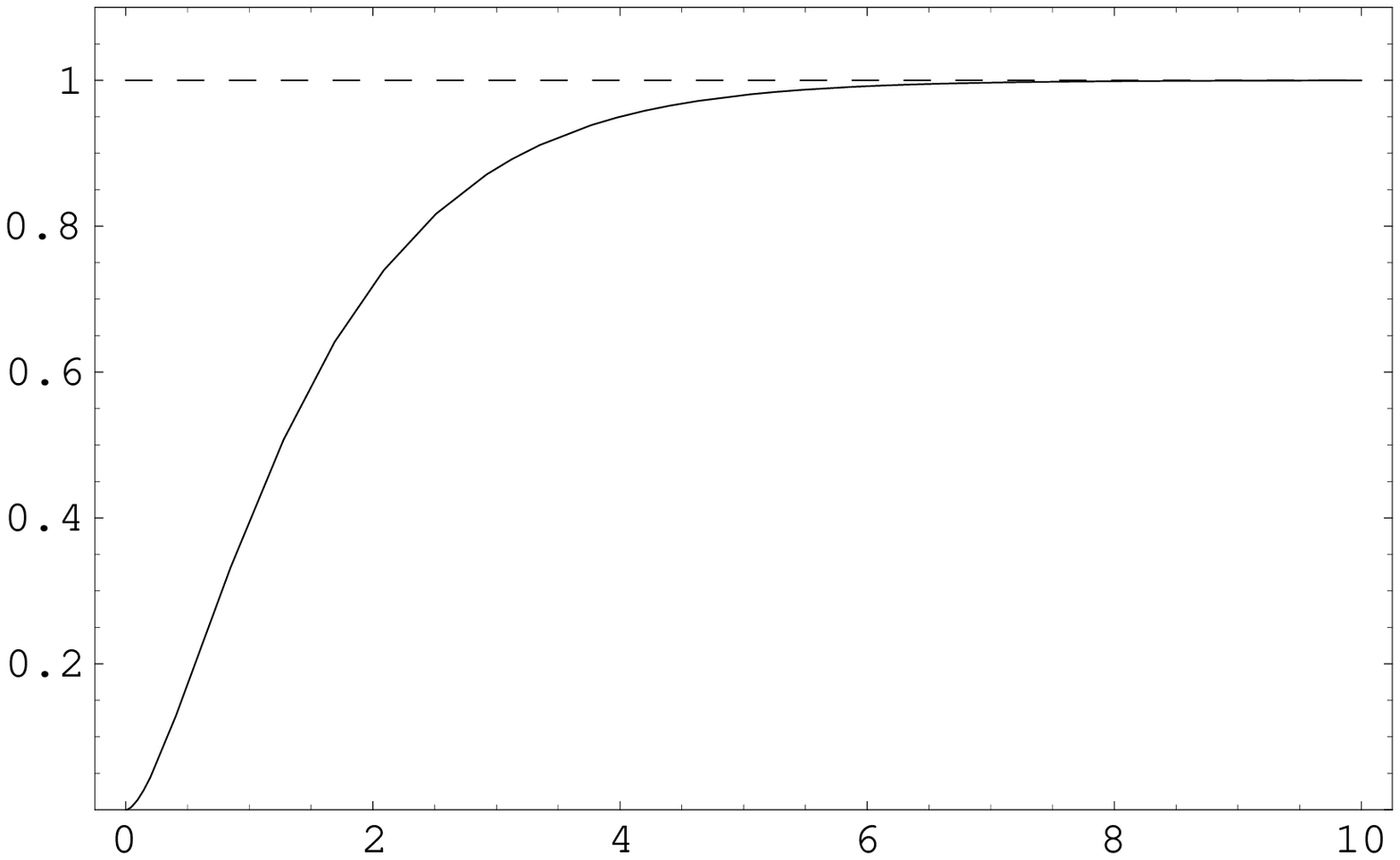,width=9cm}}
\put(4.5,0.0){$\kappa r$}
\put(-1.0,4.5){$\frac{b^z(r)}{b}$}
\end{picture}
}
\caption{Effective Burgers vector $b^z(r)/b$ (solid).}
\label{fig:burger}
\end{figure}

Let us now rewrite the distortion~(\ref{disl-pot})
as follows
\begin{align}
\label{disl-pot2}
\beta^{z}&=\frac{b}{2\pi}\Big(\d\big\{(1-\kappa r K_1(\kappa r))\varphi\big\}
-\varphi\kappa^2 rK_0(\kappa r)\,\d r\Big)\\
&\equiv \d u^z +\phi^z\nonumber,
\end{align}
where $\varphi=\arctan (y/x)$.
We can interpret the field 
\begin{align}
\phi^z=-\frac{b\kappa^2}{2\pi}\,\varphi r K_0(\kappa r)\,\d r
\end{align}
as the proper incompatible part (negative plastic distortion) 
of the distortion due to $\d\phi^z\neq 0$. 
It vanishes at $r=0$ and $r\rightarrow\infty$
(see Fig.~\ref{fig:gauge}).
\begin{figure}[h]\unitlength1cm
\centerline{
\begin{picture}(9,8)
\put(0.0,0.2){\epsfig{file=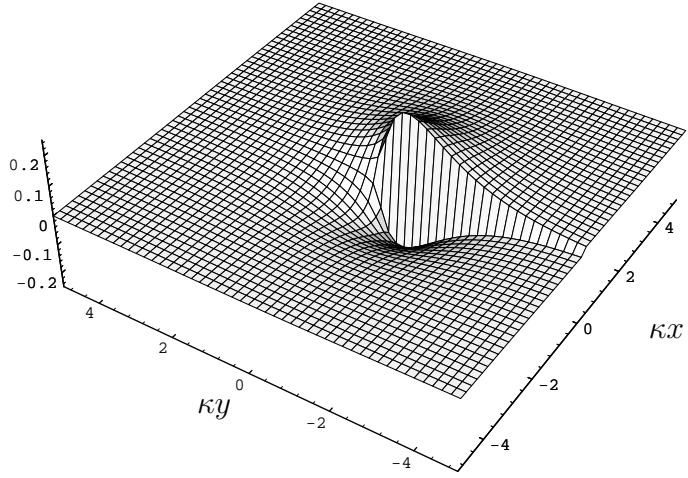,width=9cm}}
\put(8.5,2.0){$\kappa x$}
\put(2.5,1.0){$\kappa y$}
\end{picture}
}
\caption{Incompatible plastic distortion $-\phi^z /b\kappa$.}
\label{fig:gauge}
\end{figure}
The compatible part of Eq.~(\ref{disl-pot2}) reads
\begin{align}
\label{u-z}
u^z=\frac{b}{2\pi}\big(1-\kappa r K_1(\kappa r)\big)\varphi,
\end{align}
and is a modified displacement field (see Fig.~\ref{fig:u_z}). 
This $u^z$ is multivalued and has no singularity.
\begin{figure}[h]\unitlength1cm
\centerline{
\begin{picture}(9,8)
\put(0.0,0.2){\epsfig{file=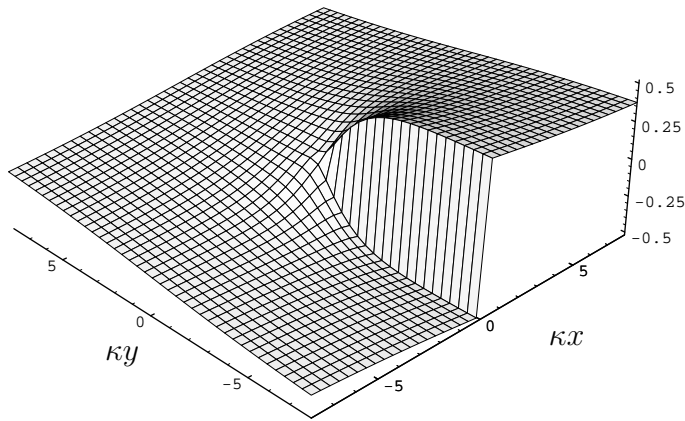,width=9cm}}
\put(7.2,1.2){$\kappa x$}
\put(1.3,1.0){$\kappa y$}
\end{picture}
}
\caption{Displacement field $u^z/b$.}
\label{fig:u_z}
\end{figure}
The asymptotic form of $u^z$ is the classical 
displacement function $\frac{b}{2\pi}\varphi$ and it vanishes at $r=0$.
Moreover, we observe that the classical displacement function is only a kind 
of {\it phase only} approximation and is not valid in the dislocation core 
analogous to the Higgs field in superconductors or in string theory.
Obviously, the displacement field $u^z$ plays the role of a 
{\it Higgs field} in the elastoplastic theory. 
The line $u^z=0$ is surrounded by a tube of the radius $\approx r_{\rm c}$,
the dislocation core, within which $b^z(r)$ is suppressed
from its constant value $b$.
From this point of view, we may identify the length, 
\begin{align}
r_{\rm c}\simeq 6/\kappa, 
\end{align}
as the {\it dislocation core radius}. 
Outside the core radius the classical elasticity describes
dislocations very well. Thus, the core radius is the inner cut-off
radius of classical elasticity where linear elasticity theory should apply.
In this framework,
Eq.~(\ref{u-z}) describes the ``atomic'' arrangement in the core region
(hopefully in good approximation).
The local atomic configuration inside the core region is fundamentally
different from that of the defect-free parts of the crystal. 
Therefore, a dislocation is 
a defect breaking locally the translation invariance in the core region.

Let us mention that Edelen and Kadi{\'c}~\cite{edelen82,edelen83} imposed the 
conditions $\xi^a=x^a$ (no elastic displacements) and $\sigma^{\rm bg}_a=0$
(no background stress) in their investigation of 
dislocation-type solutions. Accordingly, they used the translational part
of the affine connection instead of the generalized affine connection. 
That was the reason why they obtained short-reaching solutions which
are ``unphysical''.
 
After all these considerations, we may identify the mapping function 
from the defect-free to the distorted configuration according to
\begin{align}
\label{map-screw}
\xi^z=z+\frac{b}{2\pi}\big(1-\kappa r K_1(\kappa r)\big)\varphi.
\end{align}
The corresponding anholonomic coframe of the inner geometry is given by
\begin{align}
\label{coframe-screw}
\vartheta^r=\d r,\quad
\vartheta^\varphi=r\d\varphi,\quad
\vartheta^z\equiv\d z+\beta^z
=\d z+\frac{b}{2\pi}\big(1-\kappa r K_1(\kappa r)\big)\d\varphi.
\end{align}
This coframe has a helical structure and no artificial singularity.

The {\it force stress of a screw dislocation} is given by
\begin{align}
\label{stress-screw}
\sigma_{z\varphi}=\sigma_{\varphi z}	       
		= \frac{\mu b}{2\pi r}\big(1-\kappa r K_1(\kappa r)\big).
\end{align}
This eigenstress of a screw dislocation is modified near the dislocation
core (up to $6/\kappa$) and decays like $r^{-1}$ for large $r$.
It does not possess singularity at $r=0$. The eigenstress has a maximum 
at $r\simeq 1.1\kappa^{-1}$ (see~Fig.~\ref{fig:stress}):
\begin{align}
\label{stress_max}
\sigma^{\rm max}_{z\varphi}\simeq0.399\frac{\mu b}{2\pi}\kappa.
\end{align}
\begin{figure}[h]\unitlength1cm
\centerline{
\begin{picture}(9,6)
\put(0.0,0.2){\epsfig{file=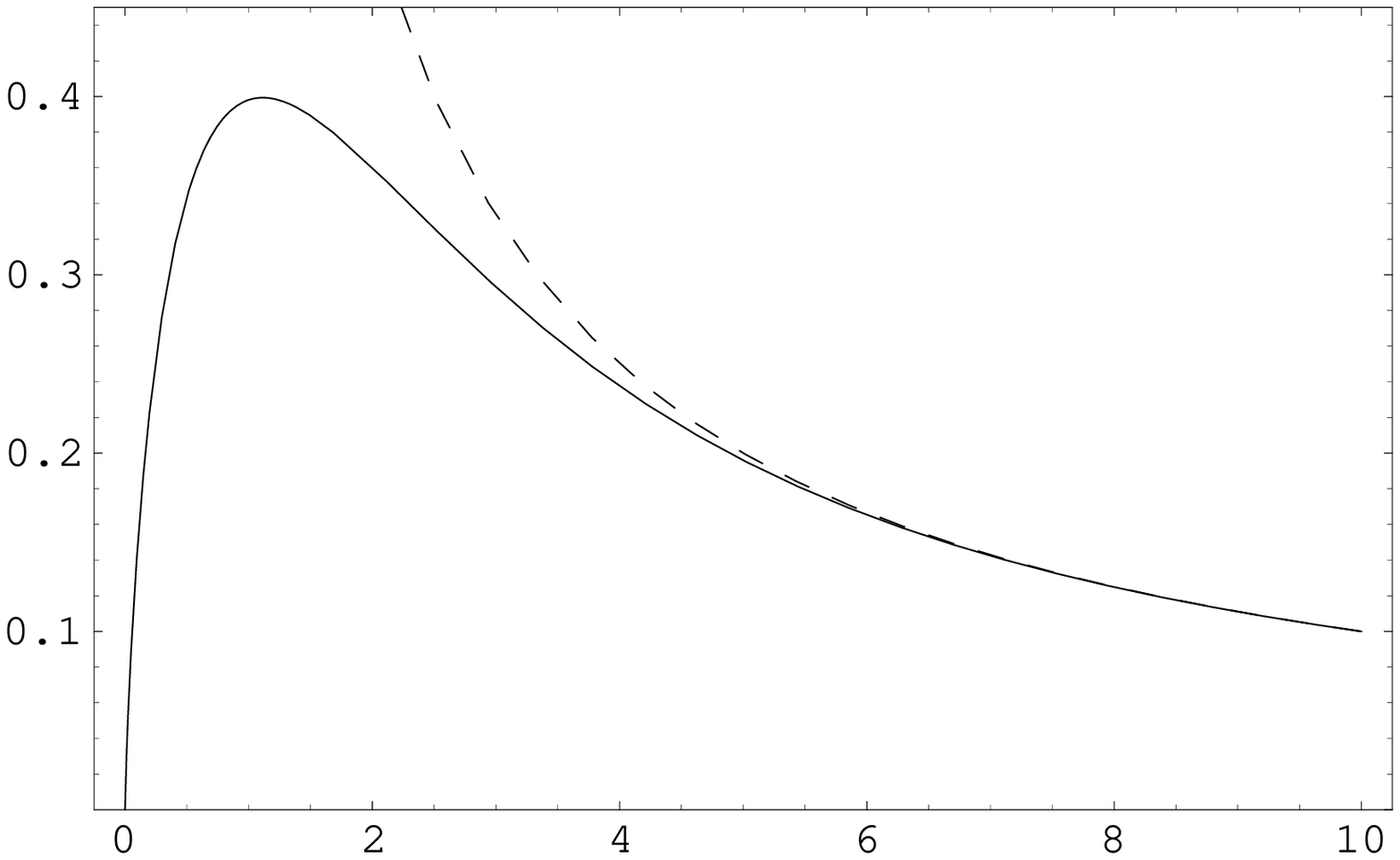,width=9cm}}
\put(4.5,0.0){$\kappa r$}
\put(-1.5,4.5){$\sigma_{z\varphi}\frac{2\pi}{\mu b\kappa}$}
\end{picture}
}
\caption{Force stress of a screw dislocation  $\sigma_{z\varphi}(2\pi/\mu b \kappa)$ (solid) and classical 
$1/r$-stress (dashed).}
\label{fig:stress}
\end{figure}
Consequently, it is not true that the eigenstress of a screw dislocation
decays exponentially with distance $r$ far from the core.  

Let us mention that the modified stress field (\ref{stress-screw}) agrees
with Eringen's stress field~\cite{eringen83,eringen90} which is 
calculated in the framework of {\it nonlocal elasticity}. 
Additionally, 
it is interesting to note that the stress field~(\ref{stress-screw})
is the same as the one obtained by Gutkin and Aifantis~\cite{gutkin99,gutkin00}
in their version of {\it gradient elasticity}.

\subsection{Torsion, moment stress, energy and force of screw dislocations}
Let us now apply the gauge potential of a screw dislocation in order 
to calculate the torsion, the moment stress, the elastoplastic energy, 
and the dislocation core energy. Additionally, we compute the modified 
Peach-Koehler force between two screw dislocations.

The nonvanishing components of torsion are now calculated by means of the 
dislocation potential similar to a magnetic vortex in cylindrical coordinates 
\begin{align}
\label{torsion}
T^z=-\frac{b\kappa}{2\pi}\frac{\pd}{\pd r}\big(r K_1(\kappa r)\big)\, \d r\wedge\d\varphi
		=\frac{b\kappa^2}{2\pi}\, r K_0(\kappa r)\,\d r\wedge\d\varphi,
\end{align}
and in Cartesian coordinates as
\begin{align}
\label{torsion-cart}
T^z=\frac{b\kappa^2}{2\pi}\, K_0(\kappa r)\,\d x\wedge\d y,
\end{align}
where $K_0$ is the modified Bessel function of the second kind of order 
zero~(see~Fig.~\ref{fig:torsion}). Let us note that this elastoplastic field
strength (torsion) is analogous to the magnetic field strength of a magnetic 
vortex (see Tab.~\ref{tab1}). 
Of course, the torsion $T^z$ fulfills the Bianchi identity $\d T^z=0$.
Note that $T^z_{\ r\varphi}\approx-\frac{b\kappa^2}{2\pi} (\ln\frac{\kappa r}{2}+\gamma)$ for
$r\ll \kappa^{-1}$ (near field) and 
$T^z_{\ r\varphi}\approx\frac{b\kappa^2}{2\sqrt{2\pi\kappa r}} \exp(-\kappa r)$ 
for $r\gg \kappa^{-1}$ (far field). Thus the far field of torsion 
decreases exponentially with $r$, with the {\it characteristic length} $\kappa^{-1}$.
\begin{figure}[h]\unitlength1cm
\centerline{
\begin{picture}(9,6)
\put(0.0,0.2){\epsfig{file=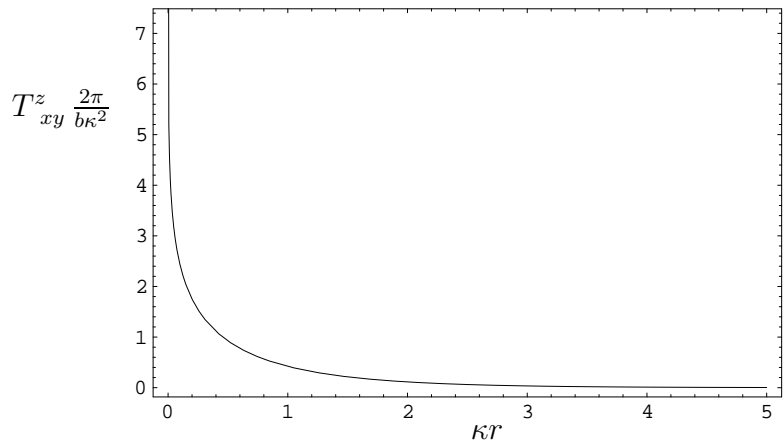,width=9cm}}
\put(4.5,0.0){$\kappa r$}
\put(-1.6,4.3){$T^z_{\ xy}\,\frac{2\pi}{b\kappa^2}$}
\end{picture}
}
\caption{Dislocation density  $T^z_{\ xy}(2\pi/b\kappa^2)$.}
\label{fig:torsion}
\end{figure}
When $\kappa^{-1}\rightarrow 0$ in (\ref{torsion}) we obtain the Dirac delta function as torsion
and dislocation density, respectively, so that the ``classical'' dislocation 
density is reverted in this limit.
Additionally, we observe that the dislocation 
density~(\ref{torsion}) agrees with Eringen's two-dimensional nonlocal 
modulus which was obtained by matching the phonon dispersion 
curves~\cite{eringen83}. 
Now, we define the {\it plastic penetration depth},
\begin{align}
R_{\rm c}:=\frac{1}{\kappa}=\sqrt{\frac{a_1}{2\mu}},
\end{align}
as the region over which 
the torsion is appreciably different from zero
and the torsional flux is confined within this region. 
Then $R_{\rm c}$ measures the proper plastic region where $\phi^a\neq 0$.
Moreover, the new constant $a_1$
is determined through $R_{\rm c}$ and $\kappa$, respectively.
The maximum of the stress is in the Peierls-Nabarro model~\cite{peierls,nabarro} 
given as $\mu/2\pi$ with $b=a$, where $a$ is the lattice parameter. 
If we compare our result~(\ref{stress_max}) with this maximum of the stress field,
it is possible to determine the unknown factor $\kappa$ as
\begin{align}
\label{kap}
\kappa^{-1}\simeq 0.399 a.
\end{align}
Therefore, by means of (\ref{kap}), the typical material constant for a screw 
dislocation is given by
\begin{align}
\label{a3}
a_1\simeq 2\mu(0.399 a)^2.
\end{align}
Let us mention that Eringen has already been obtained an analogical result
for $\kappa^{-1}$ in his nonlocal elasticity theory~\cite{eringen83}. 
He pointed out that the choice of $\kappa^{-1}\simeq 0.399 a$ excellently matches
with experimental atomic dispersion curves.
For this value of $\kappa^{-1}$ the core radius is given as
$r_{\rm c}\simeq 2.4 a$.
The stress field has its maximum $\mu b/2\pi$ at $r\simeq 0.44 a$. 
Gutkin and Aifantis~\cite{gutkin99,gutkin00} have used another choice of the 
factor $\kappa$ as $\kappa^{-1}\simeq 0.25 a$ so that the core radius 
reads $r_{\rm c}\simeq 1.50 a$.
Thus, the factor $\kappa$ determines the position and the magnitude of the 
stress and strain maximum.
Finally, the factor $\kappa$ should be fitted by comparing predictions of the
theory with experimental results and computer simulations.

The presence of dislocations gives rise to a localized moment stress.
This moment stress one-form is given by the help of Eq.~(\ref{moment5}) as
\begin{align}
H_z=
   \frac{\mu b}{2\pi}K_0(\kappa r)\,\d z,\qquad
H_x=-\frac{\mu b}{2\pi}K_0(\kappa r)\,\d x,\qquad
H_y=-\frac{\mu b}{2\pi}K_0(\kappa r)\,\d y.
\end{align}
These moment stresses mean physically twisting moments 
in the dislocation core region.
We find for the Nye tensor 
\begin{align}
\kappa_{zz}=
   \frac{b\kappa^2}{4\pi}K_0(\kappa r),\qquad
\kappa_{xx}=-\frac{b\kappa^2}{4\pi}K_0(\kappa r),\qquad
\kappa_{yy}=-\frac{b\kappa^2}{4\pi}K_0(\kappa r).
\end{align}
The Nye tensor and the moment stress are appreciable different from zero 
in the region $r\le R_{\rm c}$.

Now we are able to calculate the strain and the core energy in this framework.
The {\it stored strain energy} of a screw dislocation per unit length is given by
\begin{align}
\label{E_strain}
 E_{\rm strain}&=\frac{\mu b^2}{4\pi} \int_{0}^R\d r r\Big(\frac{1}{r}-\kappa K_1(\kappa r)\Big)^2  \nonumber\\
	       &=\frac{\mu b^2}{4\pi} 
		\Big\{\ln(r)+2K_0(\kappa r)
		+\frac{\kappa^2 r^2}{2}\big(K_1(\kappa r)^2
	        -K_0(\kappa r)K_2(\kappa r)\big)\Big\}\Big|_{0}^R,
\end{align}
where $R$ is the outer cut-off radius.
We use the limiting expressions for $r\rightarrow 0$,
\begin{align}
K_0(\kappa r)\approx-\gamma-\ln \frac{\kappa r}{2},\quad
K_1(\kappa r)\approx \frac{1}{\kappa r},\qquad
K_2(\kappa r)\approx -\frac{1}{2}+\frac{2}{(\kappa r)^2},
\end{align}
where $\gamma=0.57721566\ldots$ is the Euler constant,
and for $r\rightarrow\infty$:
\begin{align}
K_n(\kappa r)\approx \frac{\sqrt{\pi}\, \exp(-\kappa r)}{\sqrt{2\kappa r}}.
\end{align}
The final result for the strain energy reads 
\begin{align}
\label{E_strain2}
 E_{\rm strain}&=\frac{\mu b^2}{4\pi} 
		\left\{\ln\frac{\kappa R}{2}+\gamma-\frac{1}{2}\right\}.
\end{align}
Thus, we obtain a strain energy density which is not singular at the dislocation
line.
The {\it dislocation core energy} per unit length is
\begin{align}
\label{E_core}
E_{\rm core}&=\frac{\mu b^2\kappa^2}{4\pi}\int_0^\infty\d r r\, K_0(\kappa r)^2 \nonumber\\
	       &=\frac{\mu b^2\kappa^2}{8\pi} 
		r^2\left\{ K_0(\kappa r)^2-K_1(\kappa r)^2\right\}\big|_0^\infty \nonumber\\
	       &=\frac{\mu b^2}{8\pi},
\end{align}
which agrees, up to a factor 2, with  the core or misfit energy that 
is calculated for the screw dislocation in the Peierls-Nabarro 
model~\cite{HL}. 
Finally, we obtain for the total energy (per unit length) of a screw dislocation
\begin{align}
\label{E_tot}
 E_{\rm screw}=\frac{\mu b^2}{4\pi} 
		\Big\{\ln\frac{\kappa R}{2}+\gamma\Big\}.
\end{align}
Due to the fact that the Burgers vector is quantized (see section~\ref{topo}), 
the core and strain energy of a dislocation is also quantized.

Now we recover the Peach-Koehler force from Eq.~(\ref{force1}) as
\begin{align}
f^{\rm el}_a&\equiv-f^{\rm PK}_a\nonumber\\
	    &=-\pd_{[a}\beta^b_{\ j]}\sigma_b^{\ l}\epsilon_{lmn}
		\d x^j\wedge\d x^m\wedge\d x^n.	
\end{align}
For straight screw dislocations the Peach-Koehler force is given 
in the framework of linear dislocation gauge theory as a radial force density
\begin{align}
f^{\rm PK}_r
	     = \pd_r\big(\beta^{z\varphi}\sigma_{z\varphi}\big) 
		\eta.	
\end{align}
We obtain for the force per unit length acting on one screw dislocation in the 
stress field due to the other screw dislocation from Eq.~(\ref{E_strain}) 
\begin{align}
F^{\rm PK}_r&=2 \pd_r E_{\rm strain}\nonumber\\
&=\frac{\mu b^2}{2\pi r}
\Big(1-2\kappa r K_1(\kappa r)+\kappa^2 r^2 K_1(\kappa r)^2\Big).
\end{align}
Here $E_{\rm strain}$ is the interaction strain energy between the two
parallel screw dislocations.
We see that the modified Peach-Koehler force is attractive for screw dislocations of 
opposite sign, and repulsive for dislocations of the same sign and is far-reaching.
\begin{figure}[h]\unitlength1cm
\centerline{
\begin{picture}(9,6)
\put(0.0,0.2){\epsfig{file=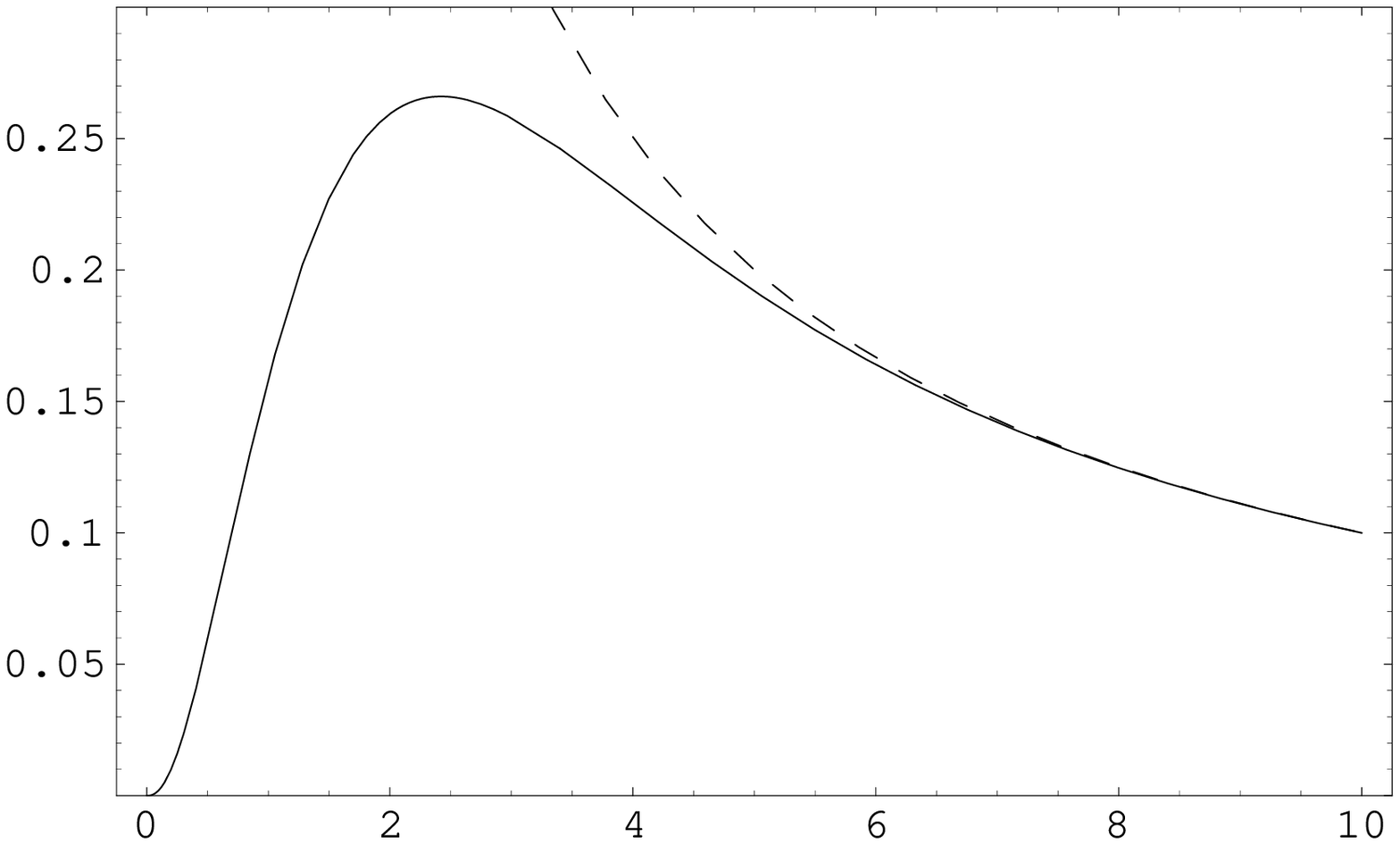,width=9cm}}
\put(4.5,0.0){$\kappa r$}
\put(-1.7,4.1){$F^{\rm PK}_{r}\frac{2\pi}{\mu b^2\kappa}$}
\end{picture}
}
\caption{Peach-Koehler force $F^{\rm PK}_{r}(2\pi/\mu b^2\kappa)$ 
(solid) and classical $1/r$-force (dashed).}
\label{fig:PK-force}
\end{figure}
This Peach-Koehler force is modified near the dislocation
core (up to $6\kappa^{-1}$) and decays like $r^{-1}$ for large $r$.
It does not possess any singularity at $r=0$. The modified Peach-Koehler force 
has a maximum at $r\simeq 2.42 \kappa^{-1}$ (see~Fig.~\ref{fig:PK-force}).

\section{Dislocation theory as three-dimensional gravity}
\setcounter{equation}{0}
In the preceding section, we have described the dislocation theory as a 
Weitzenb{\"o}ck space (teleparallelism) with nontrivial torsion $T^a$. 
An alternative description of dislocation theory is to consider the 
body manifold as a Riemann space with Christoffel symbols as a connection 
and nontrivial Riemannian curvature.
In the last case the dislocation 
theory is equivalent to three-dimensional gravity~\cite{katanaev92}.
In this picture, the Cauchy-Green strain tensor is the gravitational field 
which describes
the deformation of the manifold from the undeformed one. 

The Levi-Civita connection (Christoffel symbol) $\widetilde\omega_{ab}$,
corresponding to the metric (Cauchy-Green tensor) 
$G=\delta_{ab}\,\vartheta^a\otimes\vartheta^b$, can be derived
from the contortion one-form $\tau_{ab}$ by means of the teleparallel 
condition:
\begin{align}
\widetilde\omega_{ab}-\tau_{ab}=\omega_{ab}\equiv 0\quad
\Longrightarrow\quad
\widetilde\omega_{ab}=\tau_{ab}.
\end{align}
Then the Levi-Civita connection is given by
\begin{align}
\widetilde{\omega}_{ab}=
\frac{1}{2}\big(-T_{abi}-T_{bia}+T_{iab}\big)\d x^i .
\end{align} 
The corresponding Riemannian curvature two-form reads
\begin{align}
\label{Riem-curv}
\widetilde{R}_{ab}=\frac{1}{2}\,\widetilde{R}_{abij}\, \d x^i\wedge\d x^j
=\d\widetilde{\omega}_{ab}+\widetilde{\omega}_{ac}\wedge\widetilde{\omega}^c_{\ b}.
\end{align}
Eventually, we get the corresponding field equation 
by using the three-dimensional Hilbert-Einstein Lagrangian, 
\begin{align}
\label{Einstein-Hilbert}
{\cal L}_{\rm HE}=-\frac{1}{2\ell}\widetilde{R}^{ab}\wedge\eta_{ab},
\end{align}
instead of the teleparallel Lagrangian ${\cal L}_{\rm disl}$ in
the ``Einstein choice'' of the 
three parameters ($a_1=1$, $a_2=-1$, $a_3=-\frac{1}{2}$)
as showed in Refs.~\cite{lazar20,malyshev00}
by the help of the following remarkable identity\footnote{Note 
that $a_2=-2$ in Ref.~\cite{lazar20} is unfortunately a misprint.}
(see, e.g,~\cite{katanaev92})
\begin{align}
-R^{ab}\wedge\eta_{ab}
+\widetilde{R}^{ab}\wedge\eta_{ab}
-T^a\wedge \hodge\left(^{(1)}T_a- \,^{(2)}T_a-\frac{1}{2}\,^{(3)}T_a\right)
\equiv2\d\big(\vartheta^a\wedge \hodge T_a\big).
\end{align}
Consequently, in a Weitzenb{\"o}ck space with vanishing Riemann-Cartan 
curvature, i.e., $R^{ab}=0$, the Lagrangian ${\cal L}_{\rm disl}$ is, up 
to a boundary term, equivalent to the Hilbert-Einstein Lagrangian 
${\cal L}_{\rm HE}$ in three dimensions\footnote{The gauge theoretical description
of dislocation theory based on the Lagrangian ${\cal L}_{\rm HE}$ in combination 
with an elastic Lagrangian is also proposed by Malyshev~\cite{malyshev00} 
and is equivalent to the teleparallel formulation in this paper.}.
After variation with respect to $\vartheta^a$, one recovers an Einstein-type 
field equation
\begin{align}
\label{Einstein-eq}
\widetilde{G}_a\equiv\frac{1}{2}\,\eta_{abc}\,\widetilde{R}^{bc}=\ell\,\widehat\Sigma_a.
\end{align}
Here $\ell$ is the coupling constant of ``dislocation gravity''.

Let us analyze the Riemannian geometry caused by a screw dislocation by using
the torsion~(\ref{torsion-cart}) and the elastoplastic stress tensor.
The nonvanishing components of the Einstein tensor are 
\begin{align}
\label{Einstein-ten}
\widetilde{G}_{x}&=
-\frac{b\kappa^3}{4\pi r}\,
 K_1(\kappa r) y\,\d x\wedge\d y,\nonumber\\
\widetilde{G}_{y}&=
\frac{b\kappa^3}{4\pi r}\,
 K_1(\kappa r) x\,\d x\wedge\d y,\nonumber\\
\widetilde{G}_{z}&=
-\frac{b\kappa^3}{4\pi r}\, K_1(\kappa r)
\big(x\,\d x\wedge\d z +y\, \d y\wedge\d z\big).
\end{align}
The source of the Einstein tensor is the following effective stress tensor,
\begin{align}
\label{stress-ein}
\widehat\Sigma_{x}&=
-\frac{\mu b\kappa}{2\pi r}\,
 K_1(\kappa r) y\,\d x\wedge\d y,\nonumber\\
\widehat\Sigma_{y}&=
\frac{\mu b\kappa}{2\pi r}\,
 K_1(\kappa r) x\,\d x\wedge\d y,\nonumber\\
\widehat\Sigma_{z}&=
-\frac{\mu b\kappa}{2\pi r}\,K_1(\kappa r)
\big(x\, \d x\wedge\d z +y\, \d y\wedge\d z\big),
\end{align}
which is the eigenstress of a screw dislocation
without the ``classical'' displacement field (Higgs field) 
$u^z=\frac{b}{2\pi}\varphi$.
We see that the field $u^z$ gives no contribution to the 
Einstein tensor~(\ref{Einstein-ten}) and to the 
stress tensor~(\ref{stress-ein}). All this looks like general relativity 
where we are internal observers. One may imagine that an external observer
is able to deform our universe from outside, but this deformation would
be compatible and therefore not felt by us as internal observers\cite{kroener94}.
Is, perhaps, this observation a hint why we 
may use the Cartan or affine connection instead of the generalized
affine connection in gravity? However, the ``Higgs field'' $u^a$ should
play a physical role in gravity, too, e.g. as a nontrivial vacuum.

Obviously, the dislocation acts as the source of an incompatible 
``gravitational'' distortion field and is its own source. 
Additionally, we can say that a 
screw dislocation is a topological string with cylindrical symmetry 
in three-dimensional gravity.
Perhaps dislocations in crystals provide a better experimental field for 
testing gravity models with cosmic strings~(see also~\cite{katanaev99}).
Let me note that the interesting analogy
between vortices in superfluids and spinning cosmic strings 
is discussed in Ref.~\cite{volovik}.

Now we determine the ``gravitational'' constant $\ell$ for a screw dislocation. 
After substituting of~(\ref{Einstein-ten}) and (\ref{stress-ein}) 
in~(\ref{Einstein-eq}), we observe 
\begin{align}
\ell=\frac{1}{ a_1}.
\end{align}
We are discussing the material tungsten (W) because it is nearly isotropic.
With the lattice constant, $a=3.16\times 10^{-10}\,{\rm m}$, and 
the shear modulus, $\mu=1.61\times 10^{11}\, {\rm N}\,{\rm m}^{-2}$, we obtain 
with~(\ref{a3})
\begin{align}
\ell\simeq 1.95\times 10^{8}\, {\rm N}^{-1}.
\end{align}
Remarkable, the ``gravitational'' constant in dislocation theory 
is much bigger than the Einstein gravitational constant 
$\ell_{\rm E}=2.08\times 10^{-43}\, {\rm N}^{-1}$.
But this is not surprising because the Planck length,
$a_{\rm Pl}=1.62\times 10^{-35}\,{\rm m}$,
is much smaller than the lattice constant of the crystal.
Because the coupling constant is quadratic in the specific 
length, the difference between the gravitational constant in dislocation
theory and Einstein gravity should be a factor $10^{50}$ -- and this is what
we get.

From the quantum mechanical point of view, a crystal has a lattice structure.
We are able to measure the lattice constants by means of X-ray diffraction 
or transmission electron microscopy and we observe that a crystal is not a 
continuum. Nevertheless if the energy of the particles is so small that
the lattice structure cannot be resolved, then the differential geometric 
specification provides an effective description of the {\it continuized}
crystal (see also~\cite{froehlich}). 
A continuized crystal is the result of a limiting
process in which the lattice parameter more and more reduced such that 
the mass density and the crystallographic directions remain 
unchanged~\cite{kroener86}. 
Therefore, the dislocation theory as three dimensional gravity by means of
Einstein field equations is a low energy description like Einstein theory of
gravity. But in gravity we have not a ``microscope'' in order to observe 
the typical length and the lattice parameter, respectively.

\section{Some topological remarks about dislocations in crystals}
\label{topo}
\setcounter{equation}{0}
Let us now discuss some topological properties of dislocations in 
crystals~(see~\cite{rogula,mermin,trebin}).
Due to the Burgers circuit~(\ref{burger}), a dislocation is a topological 
line defect and the body manifold ${\cal M}^3$ is not simply connected (that is, 
if ${\cal M}^3$ contains incontractible loops). 
In general line defects, e.g., dislocation, vortices, and cosmic strings, 
are described by the first homotopy group $\pi_1$.

The three-dimensional crystal is described by the discrete translations in 
three dimensions (Bravais lattice vectors). The isotropy group of the crystal 
is the group of discrete translations ${\Bbb Z}^3$. If we identify points differing by a 
primitive lattice vector, we see that the one-dimensional translation group 
$T(1)$ is mapped to the group $U(1)$ and the one-dimensional sphere $S^1$, 
respectively. After this periodic boundary condition the corresponding space
is identified with the three-dimensional torus 
$T^3\cong T(3)/{\Bbb Z}^3=S^1\times S^1\times S^1$ and the continuous translation
group is broken to the discrete translation group: $T(3)\rightarrow {\Bbb Z}^3$. 
The type of defect depends on the topology of $T^3$.
The first fundamental group of the three-dimensional torus as the coset space 
is $\pi_1(T^3)={\Bbb Z}^3$. 
Thus, from the topological point of view, the dislocations are characterized 
by a Bravais lattice vector 
${\bf b}=u{\bf a}_1+v{\bf a}_2+w{\bf a}_3$, called
Burgers vector. Here $({\bf a}_1,{\bf a}_2,{\bf a}_3)$ are the primitive 
lattice vectors and $(u,v,w)\in\pi_1(T^3)$. 
Thus the Burgers vector is quantized. Due to the non-vanishing of the first 
homotopy group $\pi_1(T^3)={\Bbb Z}^3$, the underlying fibre bundle is topologically
non trivial.
 
We have seen that the dislocations are topological defects similar to 
vortices, where the magnetic flux 
is quantized. In general, topological defects are known as topological charges
in gauge theories. 
The quantized abelian topological charge of dislocations is the Burgers vector 
that is the torsional flux. Hence, a dislocation is a kind of a torsion vortex in crystals.
Dislocations have (pseudo) particle-like properties. For example, they may 
annihilate with their ``anti-particles'', i.e., dislocations of opposite 
Burgers vector. Let us remark, that Seeger~\cite{seeger80} has already been 
considered dislocations as solitons in crystals, namely, 
as global soliton or kink which is a solution of the Enneper or sine-Gordon
equation.

Some topological remarks about static dislocations have been
discussed by Gairola~\cite{gairola93}. But he has not clarified the nature of 
the gauge field, which we have identified with the dislocation gauge
potential in the framework of translation gauge theory~\cite{lazar20}.

Dislocations in crystals can be described by means of the Burgers vector and 
the direction of the dislocation line ${\bf s}$. One usually distinguishes 
between screw (${\bf b}\|{\bf s}$) and edge (${\bf b}\perp{\bf s})$ dislocations. 
But from the topological point of view these both types are equivalent.

\section{Conclusions}
We have proposed a static theory of dislocations with moment stress 
which represents the specific response to dislocation distributions
in an anisotropic or isotropic elastoplastical material as a three-dimensional 
translation gauge theory. 
We have explicitely been seen that a physical field theory of dislocations
has to contain the notion of moment stress. 
Hence, dislocation theory is a couple or moment stress theory.
In this theory of dislocations the force stress vanishes except at the positions
of the dislocations, where it gives rise to a localized moment stress. 
Obviously, the size of this moment stress cannot be calculated from classical
elasticity theory.
Thus, a field theory of dislocations without moment stress is obsolete.  

In our theory we have used the framework of MAG and the analogy
between the dislocation theory and Maxwell's theory. In order to 
obtain a field theory,
we have used the concepts of field strength, excitation, and constitutive law
analogical to the electromagnetic field theory. 
All elastoplastical field quantities can be described by ${\Bbb R}^3$-valued
exterior differential forms. The elastoplastic field strength is
an even (or polar) differential form and the moment and the force
stress are odd (or axial) forms.
We have shown that the elastoplastic excitation with respect to 
dislocation density is necessary for a 
realistic physical dislocations theory.
As constitutive relation between dislocation density and moment stress 
we have discussed linear laws for isotropic and anisotropic materials.
For isotropic materials we used  the teleparallel Lagrangian, which
is equivalent to the Hilbert-Einstein Lagrangian, as dislocation gauge 
Lagrangian. In this case, the constitutive relation between the dislocation 
density and the moment stress is compatible with the constitutive law between
the strain and the symmetrical force stress. 
Moreover, we have proven that the moment stress in the ``Einstein-choice'' is
proportional to the Nye tensor.
A new material constant $a_1$ enters in the constitutive relation between 
the dislocation density and the moment stress. It defines a new 
internal length scale $\kappa^{-1}$.

Additionally, we have demonstrated how to fit the excitations into the 
Maxwell type field equations in contrast to Ref.~\cite{goleb} who claimed 
that there are no analogues to the second pair of Maxwell equations in 
dislocation theory. 
A static dislocation theory is analogous to the magnetostatics.
We have used the analogy between fields which have the same field theoretical
meaning (differential forms of the same degree). 
Therefore, from the field theoretical point of view, this analogy is more
straightforward than the analogy used by Kr{\"o}ner~\cite{kroener55}.
Moreover, we have pointed out
the analogy between a magnetic (Abrikosov-Nielsen-Olesen) vortex and 
a screw dislocation in a crystal. Consequently, a dislocation is
a translational vortex or string.
A review of the corresponding magnetic and dislocation quantities 
is given in Table~\ref{tab1}.
\begin{table}[h]
\caption{The correspondence between a magnetic vortex and a screw dislocation}
\begin{tabular}{ll}
\label{tab1}
$B$ - magnetic field strength & $T^a$ - dislocation density  \\
$H$ - magnetic excitation     & $H_a$ - moment stress \\
$A$ - magnetic potential        & $\vartheta^a$ - incompatible distortion\\
$f$ - Higgs field      & $\xi^a$ - mapping function\\ 
$B=\d A$,\quad $A=A^\prime+\d f$ & $T^a=\d\vartheta^a$,\quad $\vartheta^a=\phi^a+\d\xi^a$\\
Coulomb gauge: & Coulomb gauge:\\
$\d\hodge A=0$ & $\d\hodge\vartheta^a=0$\\
magnetic flux: ($n$-winding number) & Burgers vector:\\
$\Phi_0= n \pi\hbar c/e_0$   & $b^a$ must be a lattice vector \\
$\oint_{\gamma} A=\Phi_0$ & $\oint_{\gamma} \vartheta^a=b^a$\\
$j$ - electric current & $\Sigma^{\rm T}_a=\widehat\Sigma_a+h_a$ - force stress \\
gauge potential of a magnetic string: ($n=1$)& distortion of a screw dislocation:\\
$A=\Phi_0/(2\pi)(1- \lambda rK_1(\lambda r))\d\varphi$ & $\beta^z=b/(2\pi)(1-\kappa r K_1(\kappa r))\d\varphi$\\
field strength of a magnetic string: ($n=1$)& torsion of a screw dislocation:\\
$B=\Phi_0\lambda^2/(2\pi)\, r K_0(\lambda r)\d r\wedge\d\varphi$ & $T^z=b\kappa^2/(2\pi)\, rK_0(\kappa r)\d r\wedge\d\varphi$\\
magnetic field closed: & dislocation density closed:\\
$\d B=0$        & $\d T^a=0$\\
(static) Oersted-Amp{\`e}re law: & moment stress equilibrium:\\
$\d H=j$ & $\d H_a=\Sigma^{\rm T}_a$\\
continuity equation :& force stress equilibrium:\\
$\d\, j=0$ & $\d\Sigma^{\rm T}_a=0$\\
constitutive law:& constitutive law:\\
$H=H(B)$ & $H_a=H_a(T^b)$\\
magnetic energy density: & energy density of dislocations: \\
${\cal E}_{\rm em}=\frac{1}{2}B\wedge H$ 
& ${\cal E}_{\rm disl}=\frac{1}{2}T^a\wedge H_a$\\
\end{tabular}
\end{table}

Additionally, we discussed the dislocation theory as a gravity theory in 
three-dimensions. We pointed out some similarities between dislocations and 
cosmic strings.

In classical theory of dislocations one usually claims that the dislocation
core cannot be described in linear approximation because of the singularity of
the stress field at $r=0$ and that one has to use the nonlinear elasticity
near the core.
The reason for this assumption is that the classical theory of dislocations
does not use a constitutive law between dislocation density as 
elastoplastical field strength and the moment stress as elastoplastic 
excitation in field theoretical way. In the elastoplastic field theory, it
is possible to describe the core region even in linear approximation 
very well.

Two characteristic distances appear naturally in this approach: 
the dislocation core radius $r_{\rm c}\simeq 6 \kappa^{-1}$ and 
the plastic penetration depth $R_{\rm c}\simeq  \kappa^{-1}$
which may be viewed as the radius of the region over which the dislocation
density (torsion), the Nye tensor and the moment stress are appreciably 
different from zero.
We found in this theory of dislocations with moment stress 
that the near stress field for a screw dislocation 
is modified up to $r_{\rm c}\simeq 6 \kappa^{-1}$ (core radius) and 
the far field is in agreement with the classical stress field. 
Thus the translation gauge theory of dislocations removes the artificial 
singularity at the core in the classical dislocation theory.
It gives 
the correct results of the elasticity theory for a screw dislocation and
the modification in the core region due to the moment stress.
We have discussed the choice of the coupling constant between the dislocation 
density and the moment stress  
as $a_1\simeq 2\mu (0.399 a)^2$ for a screw dislocation.
Accordingly, we obtained that the Burgers vector
is also  modified in the region from $0\le r\lesssim 2.4a$. 
Moreover, we calculated the dislocation density, the moment stress
and the elastoplastic energy of a screw dislocation. We have shown that 
this translational gauge model is useful in determining 
the width of a screw dislocation and in 
estimating the core energy of a screw dislocation similar to the Peierls-Nabarro
model.

Last by not least, we have seen that the translational gauge theory of 
dislocations is a field theory where the torsion and the translational part 
of the generalized affine connection play a physical role.
 
\acknowledgments
The author wishes to thank Profs. Friedrich W.~Hehl, Luciano Mistura,
and Alfred Seeger and Drs. Gerald Wagner and Michael Zaiser
for many stimulating and helpful discussions, furthermore to Dr. Cyril Malyshev 
for correspondence and useful comments on an earlier version of this paper.
He would like to express his gratitude to the Max-Planck-Institut f{\"u}r Metallforschung 
and to the Institut f{\"u}r Theoretische und Angewandte Physik,
University of Stuttgart  for their support during his stay in Stuttgart.

Additionally, I thank Prof. Ekkehart Kr{\"o}ner for a series of 
clarifying discussions about dislocation theory. Unfortunately, he passed 
away on 19 April 2000. The theory of dislocations owes much to Prof. E.~Kr{\"o}ner.
He will not be forgotten.


\end{document}